\documentclass[twocolumn,floats,pra,showpacs]{revtex4-1}

\usepackage{amsmath}
\usepackage{amsfonts}

\usepackage{tikz}

\newcommand {\be}{\begin{equation}}
\newcommand {\ee} {\end{equation}}
\newcommand {\bea}{\begin{eqnarray}}
\newcommand {\eea} {\end{eqnarray}}
\newcommand{\non}{\nonumber}

\newcommand{\cepn}{{\cal EP}_n}

\begin{document}

%@twocolumnfalse\endcsname

\title{Non-Hermitian adiabatic transport in spaces of exceptional points}
\author{J. H\"oller$^1$, N. Read$^{1,2}$, and J.G.E. Harris$^{1,2}$}
\affiliation{$^1$Department of Physics, Yale University, P.O. Box 208120, New Haven, CT 06520-8120\\
$^2$Department of Applied Physics, Yale University, P.O. Box 208284, New Haven, CT 06520-8284}
\date{August 10, 2020}

\begin{abstract}
We consider the space of $n \times n$ non-Hermitian Hamiltonians ($n=2$, 
$3$, \ldots) that are equivalent to a single $n\times n$ Jordan block. We focus on adiabatic transport around a 
closed path (i.e.\ a loop) {\em within} this space, in the limit as the time-scale $T=1/\varepsilon$ taken to traverse the loop
tends to infinity. We show that, for a certain class of loops and a choice of
initial state, the state returns to itself and acquires a complex phase that is $\varepsilon^{-1}$ times an 
expansion in powers of
$\varepsilon^{1/n}$. The exponential of the term of $n$th order (which is equivalent to
the ``geometric'' or Berry phase modulo $2\pi$), is thus independent of $\varepsilon$
as $\varepsilon\to0$; it depends only on the homotopy class of the loop and is an integer power of $e^{2\pi
i/n}$. One of the conditions under which these results hold is that the state being transported is, for all points
on the loop, that of slowest decay.
\end{abstract}

%\pacs{pacs} %]

\maketitle

%%%%%%%%%%%%%%%%%%%%%%%%%%%%%%%%%%%%%%%%%%%

\section{Introduction}
\label{intro}

%%%%%%%%%%%%%%%%%%%%%%%%%%

\subsection{Background}

The introduction of weak linear dissipation or amplification into a system of $n$ classical harmonic
oscillators results in time evolution that can be described using an $n \times n$ Hamiltonian matrix $H$
that is non-Hermitian. The non-Hermiticity of $H$ gives rise to the familiar decay (or growth) of such
a system's
eigenstates (normal modes). It also opens the possibility of ``exceptional points'' (EPs) in
parameter space, at which $H$ is not fully diagonalizable. In the neighborhood of an EP, the
eigenvalues exhibit branch-point behavior as functions of the parameters, and so encircling it permutes
the eigenvalues and eigenspaces \cite{kato,miniatura,heiss}, an effect referred to as
flipping, monodromy, or spectral flow.

In recent years, EPs have been studied experimentally in a wide range of settings, including
microwave \cite{dghhhrr,dmbkglmrmr}, electrical \cite{shs}, optical \cite{lymlskla}, cavity QED
\cite{cklkkla}, exciton \cite{geblfbkshynktdo}, acoustic \cite{fsa}, and mechanical \cite{harris}
systems. While each
of these realizations has offered some degree of control over $H$, in most experiments the number $m$ of
independent control parameters is insufficient to specify an arbitrary $H$. As a result, EPs are typically
observed to occur at isolated points within the $m$-dimensional space of control parameters. In contrast,
if we consider the space $M_n({\bf C})\cong {\bf C}^{n^2}$ of all $n\times n$ complex matrices $H$, then EPs
are not isolated, but in fact form subspaces of $M_n({\bf C})$ of dimension larger than zero
\cite{miniatura,Xu2018}. These subspaces are topological spaces (not vector spaces), and in the 
neighborhood of a generic point in such a subspace it is a smooth (indeed, complex analytic) manifold;
we usually refer to these subspaces simply as spaces of EPs. 
One part of the following work is the description of the geometry and topology of these spaces 
in the simplest cases; we also explain how this is relevant to the topic of adiabatic evolution that 
we wish to study, and to which we now turn.

Leaving aside EPs for a moment, the evolution of a system under an asymptotically slow (``adiabatic'') smooth 
variation of some parameters in time has been the subject of much study in both Hermitian and non-Hermitian 
cases. In what follows, we concentrate on evolution along a closed path (a loop) in parameter space 
[for example, $M_n({\bf C})$].
In Hermitian systems, the adiabatic theorem \cite{bf,kato_adiabatic} guarantees, in terms of the
eigenstates of the ``instantaneous'' Hamiltonian at any point of the loop, that if the system is initially
in an eigenstate (or in a subspace in Hilbert space of degenerate eigenvalues) and if the degeneracy 
of that eigenvalue does not
change at any point during the evolution, then at the end of the adiabatic evolution the system will be
found in the same eigenspace in which it started. Moreover, the phase of the state vector
changes by an amount whose asymptotic form, as the time $T$ taken for the loop tends to infinity, has two
leading contributions: the integral of the eigenvalue along the loop (the dynamical phase), which
typically is linear in $T$; and the geometric or Berry phase, which is independent of $T$ \cite{berry}.
The Berry phase modulo $2\pi$, or phase factor, is the holonomy of a natural connection (i.e.\ ``vector
potential''), and may have
further topological significance \cite{berry,bs}. (For evolution of a degenerate eigenspace, the phase
factor becomes a unitary map \cite{wz}.)

In contrast, in non-Hermitian systems there exists a mode or modes that are
``dominant'', meaning they have the largest rate of exponential growth (or the slowest decay). During
adiabatic evolution along a generic loop
the system tends to transition into one of these modes, which then dominates at long times, even when the
eigenvalue (or its real or imaginary parts) of the chosen mode does not coincide with that of another
mode at any point on the loop. In this situation the adiabatic theorem does not hold for all the
modes, but only for the dominant one \cite{nenciu}. If a different mode becomes dominant somewhere
along the loop, then even this statement breaks down. This occurs generically (because of monodromy, 
which was mentioned already) when the loop {\em encircles} an EP without passing through one; as a 
result, no adiabatic theorem applies for strict adiabatic transport around such a loop \cite{uzdin,bu}. 
We note that some of the experimental literature (e.g.\ Refs.\ \cite{dghhhrr,dmbkglmrmr,harris}) 
is concerned instead with a ``quasi-adiabatic'' limit 
of evolution that is slow, but not asymptotically slow. In that regime, after evolution along a loop 
that encircles an EP, the system may end in a mode different from the one in which it began 
\cite{uzdin,milburn}, 
demonstrating the monodromy around the EP. We emphasize that this is not the same as the strict 
(i.e.\ asymptotically slow) adiabatic evolution that we consider in this paper.

%%%%%%%%%%%%%%%%%%%%%%%

\subsection{Overview of results}
\label{outline}

In this paper, instead of adiabatic transport along a loop that never passes through any EP, we consider 
loops that lie entirely {\em within} one of the spaces of EPs that were mentioned already. The simplest 
example concerns a $2\times 2$ non-Hermitian Hamiltonian matrix $H$; its EPs occur when $H$ is similar 
to a single $2\times 2$ Jordan block. [We recall the Jordan canonical form: an arbitrary matrix can be 
transformed by change of basis (i.e.\ similarity transformation) into block diagonal form, where the 
diagonal blocks are Jordan blocks, and the other blocks are zero; in each Jordan block, the diagonal 
elements are all the same eigenvalue. The vectors in such a basis are termed generalized eigenvectors. 
The Jordan canonical form of the matrix is unique up to permutations 
of the blocks.] The eigenvalue on the diagonal in the Jordan block can be set to zero by adding a multiple 
of the identity; throughout our discussion, we will assume this is already done, as allowing it to be nonzero 
produces only very simple changes. 

These EPs form a single space that we call $\mathcal{EP}_2$, which is a subspace 
of $SM_n({\bf C})\subset M_n({\bf C})$, the space of traceless complex $n\times n$ matrices, here with $n=2$ 
(we give details below); traceless $2\times 2$
Hamiltonians not in $\mathcal{EP}_2$ are fully diagonalizable. This example has simple geometry 
and is quite tractable. As it can be readily generalized to the case of traceless $n\times n$ Hamiltonians similar 
to a single $n\times n$ Jordan block $J_n$,
\be
J_n=\left(\begin{array}{ccccc}0&1&&&\\
                            &0&1&&\\
                            &&&\ddots&\\
                            &&&&1\\
                            &&&&0
                            \end{array}\right),
\ee 
and with similar results for all $n>1$, we carry out the analysis in this more general case. 
For each $n$, we denote the space of these EPs by $\mathcal{EP}_n\subset SM_n({\bf C})$. For $n>2$, 
there are also EPs in $SM_n({\bf C})$, but not in $\cepn$, at which other Jordan block structures arise, and 
the situation becomes much more complex. We do not consider those cases in the present paper.

We first describe the geometry of the space $\cepn$, with particular emphasis on the simplest
case, $n=2$. We find in particular that these spaces are $n$-fold connected (doubly connected for $n=2$),
in the sense that there are closed paths (loops) lying in $\cepn$ that cannot be contracted 
(within $\cepn$) to a point, but traversing such a loop $n$ times produces a loop that can be so contracted. 
In other words, one may associate a winding number (defined modulo $n$) to any loop, and concatenation of
two loops that both begin and end at the same point gives a loop whose winding number is the sum of those 
of the two given loops, modulo $n$. 

Then, as for the usual adiabatic theorems, we consider a loop
in $\cepn$, parametrized by Hamiltonians $H(s)$ for $0\leq s\leq 1$, where $H(1)=H(0)$, and evolve the
system in time $t$ with $s=t/T$; finally the asymptotics as $T\to\infty$ are studied, with the loop
$H(s)$ fixed (independent of $T$). We find that the result of adiabatic transport in $\cepn$ has features
in common with the case of a non-degenerate eigenstate in a Hermitian system, but also substantial
differences.
For a class of loops and a choice of initial state vector (to be described in a moment), we find that as
$T\to\infty$ the state vector returns to itself, multiplied by a complex number whose logarithm
has the Puiseux series form
\be
\sum_{r=1}^n T^{1-r/n}\int_0^1 ds\, a_r(s)
\label{puis_summ}
\ee
(plus terms higher order in $1/T^{1/n}$), where $a_r(s)$ are complex functions that can be calculated
from $H(s)$. Thus the (complex) dynamical phase (the terms with $r<n$) includes fractional
powers of $T^{-1}$ (of which the $r=1$ term at least has been noticed previously \cite{joye}). The term
of order $T^0$ (the $r=n$ term) is the geometric or Berry phase, and is only well-defined modulo $2\pi i$.
Remarkably, the exponential of this term is again the holonomy of a connection, and is precisely 
$e^{2\pi i/n}$ raised to the power of the winding number of the loop in $\cepn$; it is invariant 
under small deformations of the loop within that space. 

To describe the conditions under which this result holds, it is useful to change basis in the evolution
equation to the basis of instantaneous generalized eigenvectors of $H(s)$. In this basis, the effective
Hamiltonian describing evolution is $H'=J_n+T^{-1}A(s)$, where
$A(s)$ is the adiabatic (Berry) connection matrix along the loop at $s$ (details appear below). The most
generic case is that in which the lower-left element $A_{n1}$ of $A$ is nonzero, and then $H'$ is diagonalizable.
For adiabatic transport in $\cepn$, it is $H'$ (rather than $H$) that determines the dominant mode. We
assume that the same mode remains dominant everywhere on the loop, which is ensured if $A_{n1}$
does not touch the negative real axis or zero. Our result holds for such loops and when the initial state
vector is this dominant mode.

In addition, we identify an important subclass of such loops for which adiabatic transport of {\em any}
eigenstate of $J_n+T^{-1}A$ stays in that state for all time. This consists of loops for which $A(s)=A$ is
independent of $s$. As long as $A_{n1}$ is not zero, such a ``straight'' loop in
$\cepn$ produces the form (\ref{puis_summ}) for each eigenstate, not only for the dominant one (the
coefficients $a_r$ for $r\neq n$ differ for each eigenstate, however).  Loops of this form that 
possess non-zero winding number exist in $\cepn$, and the holonomy (Berry phase factor) of such 
a loop is independent of which eigenstate of $J_n+T^{-1}A$ is transported. 

We also explain how the effects can be studied experimentally for $n=2$ in two implementations, 
including one described in Ref.\ \cite{Xu2018}.

In Sec.\ \ref{deriv}, all the results just outlined are derived step by step, except for some parts that can 
be skipped without significant loss of understanding and which are relegated to the Appendices. Particular 
emphasis is placed on the simplest case, $n=2$, which we use to give examples. In Sec.\ \ref{expt}, we 
describe the experimental implementations.
Sec.\ \ref{conclusion} is the Conclusion.

%%%%%%%%%%%%%%%%%%%%%%%%%%%%%%%%%%%%%%%%%%%%%%

\section{Derivations of results}
\label{deriv}

\subsection{Geometry of $\cepn$}
\label{geom}

To obtain the results outlined in Sec.\ \ref{outline}, it is convenient to use the evolution (Schr\"odinger) 
equation in the form
$\partial_t|\psi\rangle=H|\psi\rangle$, so $iH$ would be the usual Hamiltonian of quantum mechanics. We
treat $H$ as a matrix, and $|\psi\rangle$ stands for a column vector.
Adding a multiple of the identity $I$ to a Hamiltonian merely shifts all the eigenvalues by the same
amount, so we can assume that the trace of $H$ is zero. An $n\times n$ complex Hamiltonian that is similar
to a single Jordan block with eigenvalue zero, meaning that $H=\Lambda J_n \Lambda^{-1}$
where $\Lambda$ is an invertible complex invertible matrix, is said to lie at an EP of type $EP_n$. (Note 
that the columns of $\Lambda$ are generalized eigenvectors of $H$; we denote them by $|u_i\rangle$ for $i=1$, 
\ldots, $n$.) 

To describe the geometry of the space $\cepn$ of all such $H$, first notice that the
matrices that commute with $J_n$ have the form $aI+b_1J_n+\cdots+b_{n-1}J_n^{n-1}$, where $I$ is the identity,
and $a$, $b_i$ are complex numbers. Such matrices with $a\neq 0$ form a Lie group ${\bf C}^\times\times
{\cal J}_n$, which is a subgroup of $GL_n({\bf C})$, the group of invertible $n\times n$ complex matrices.
Here ${\bf C}^\times$ is the group of nonzero complex numbers (under multiplication), and ${\cal J}_n$ is the
group of $n\times n$ matrices of the form $I+b_1' J_n+\cdots +b_{n-1}'J_n^{n-1}$, where $b_i'$ are complex 
numbers. Then $\cepn$ can be identified as the quotient space $\cepn\cong GL_n({\bf C})/[{\bf C}^\times\times 
{\cal J}_n]$; see Appendix \ref{app:geom} for further details. It is a non-compact space
of complex dimension $n(n-1)$, but has a ``deformation retract'' \cite{hatcher} onto $SU(n)/{\bf Z}_n$,
that is the group $SU(n)$ of unitary
matrices of determinant $1$, modulo its center ${\bf Z}_n$, the cyclic group of order $n$. The fundamental
group of this space is $\pi_1(\cepn)\cong{\bf Z}_n$ (see Appendix \ref{app:geom}). That means that loops 
in the space can be 
characterized (modulo small deformations) by a single winding number defined modulo $n$, as described 
in Sec.\ \ref{outline}. We note that if we allowed the single eigenvalue of $H$ to be nonzero instead of requiring
it to be zero, then the space of such Hamiltonians would be $\cong \cepn \times {\bf C}$ and 
have complex dimension higher by $1$, but the fundamental group would be unchanged. 

%%%%%%%%%%%%%%

\subsubsection{Example: $n=2$}

As illustration, for $n=2$, we can more explicitly describe traceless Hamiltonians as
\be
H=\left(\begin{array}{cc}Z&X-iY\\
                         X+iY&-Z
                            \end{array}\right), \label{twoham}
                            \ee
where $X$, $Y$, $Z$ are complex numbers. With our conventions, $iH$ would be Hermitian (with respect to
the standard inner product) if $X$, $Y$, $Z$ were all imaginary. However, for general non-Hermitian $H$,
an inner product plays no essential role, and we avoid using one on ${\bf C}^n$ at any stage. For
$n=2$, clearly there are no exceptional points other than those in ${\cal EP}_2$. If ${\bf
X}=(X,Y,Z)^T$ (the superscript $T$ denotes transpose), then $H$ is in ${\cal EP}_2$ if and only if 
$|{\rm Re}\, {\bf X}|=|{\rm Im}\, {\bf X}|>0$
and ${\rm Re}\,{\bf X}\cdot{\rm Im}\, {\bf X}=0$ \cite{miniatura} (here the standard inner product and
norm on ${\bf R}^3$ were used). If we fix $|{\rm Re}\,{\bf X}|$ to $1$, then because an ordered pair 
of orthogonal unit vectors in ${\bf R}^3$ (such as ${\rm Re}\, {\bf X}$, ${\rm Im}\,{\bf X}$) determines 
an orthonormal basis with positive orientation in ${\bf R}^3$, the space
of such pairs forms the special orthogonal group in three dimensions, or real projective $3$-space,
$SO(3)\cong {\bf RP}^3\cong SU(2)/{\bf Z}_2$. It is well-known that the fundamental group of this space
is $\pi_1\cong {\bf Z}_2$ (i.e.\ it is doubly connected; see Sec.\ \ref{outline}) \cite{hatcher}. Then 
${\cal EP}_2\cong SO(3)\times {\bf R}$, where the second factor represents $\ln|{\rm Re}\,{\bf X}|$ 
and is contractible. Hence ${\cal EP}_2$ is doubly connected also, that is $\pi_1({\cal EP}_2)={\bf Z}_2$. 

%%%%%%%%%%%%%%%%%%%%%%%%%%%%%

\subsection{Adiabatic transport in $\cepn$}
\label{adia_cepn}

\subsubsection{General statements}

Now we consider adiabatic transport in $\cepn$ for general $n$. We choose a smooth loop in $\cepn$,
so we have $H=H(s)$, a smooth function of $s\in [0,1]$ with $H(1)=H(0)$ and $H\in\cepn$ for all $s$. We evolve the
system in time $t$ from $0$ to $T>0$ with the time-dependent Hamiltonian $H=H(s=t/T)$ as in the usual
adiabatic evolution. If we express the evolution equation $\partial_t|\psi\rangle=H|\psi\rangle$ in a
basis of generalized eigenvectors $|u_i(s)\rangle$ of $H(s)=\Lambda(s)J_n\Lambda(s)^{-1}$ at each $s$
(that varies smoothly with $s$) then, for the column vector $|u\rangle=\Lambda^{-1}|\psi\rangle$ of 
components in this basis,  it takes the form
\be
\varepsilon\partial_s |u\rangle = (J_n+\varepsilon A) |u\rangle,
\label{jplusepsa}
\ee
where $\varepsilon=1/T$ and $A_{ij}=-\langle u_i|\partial_s u_j\rangle$ (i.e.\
$A=-\Lambda^{-1}\partial_s\Lambda$) is the Berry connection evaluated on the tangent vector to the loop.
Here the bras $\langle u_i(s)|$ are a smooth basis
set of row vectors dual to the basis of kets $|u_i(s)\rangle$, so
$\langle u_i(s)|u_j(s)\rangle=\delta_{ij}$ for each $s$ (this is {\em not} a use of an inner product);
they are rows of $\Lambda^{-1}$. To keep later arguments
simpler, we assume without loss of generality
that $\Lambda(s)$ is periodic [$\Lambda(1)=\Lambda(0)$], and so also
$A(1)=A(0)$. There is a residual gauge freedom when we obtain eq.\ (\ref{jplusepsa}): the form is preserved
under a further differentiable periodic $s$-dependent change of basis by $\widetilde{\Lambda}(s)\in {\bf
C}^\times\times {\cal J}_n$ for all $s$ (see Appendix  \ref{app:gauge}).

The key point now is that while $J_n$ is not diagonalizable, $J_n+\varepsilon A$ often is. As
$\varepsilon\to0$, $\det (J_n+\varepsilon A)=(-1)^{n-1}\varepsilon A_0+{\cal O}(\varepsilon^2)$, where we
write $A_0=A_{n1}(s)$ and we assume henceforth that $A_0$ is nonzero for all $s$. From the characteristic
equation, we find that the $s$-dependent eigenvalues of $J_n+\varepsilon A$  are $\lambda_\mu=
\zeta^\mu(\varepsilon A_0)^{1/n}$ ($\mu=0$, $1$, \ldots, $n-1$) to leading order in $\varepsilon$, where 
$\zeta=e^{2\pi i/n}$ (see Appendix \ref{app:evalscal}). Here we choose one $n$th root
of $A_0$, which we take to be the principal branch, for which $\arg A_0^{1/n}\in (-\pi/n,\pi/n]$, and
denote it $A_0^{1/n}$, and $(\varepsilon A_0)^{1/n}=\varepsilon^{1/n}A_0^{1/n}$ ($\varepsilon^{1/n}>0$).
Solving iteratively for each eigenvalue, we can obtain series expansions
\be
\lambda_\mu=\sum_{r=1}^\infty a_r\varepsilon^{r/n} \zeta^{\mu r}
\ee
with nonzero radius of convergence; such an expansion is called a Puiseux expansion. Note that here the
coefficients $a_r=a_r(s)$ ($a_1=A_0^{1/n}$) are independent of $\mu$ (because $\zeta^{\mu r}$ has been
extracted), because if such an expansion satisfies the characteristic equation for one value $\mu$, then
it does so for all $\mu$.
Then as $\sum_\mu \lambda_\mu = {\rm tr}\, (J_n+\varepsilon A)=\varepsilon\, {\rm tr}\, A$, we find that
\be
a_n = \frac{1}{n}{\rm tr}\, A
\ee
and $a_{2n}=a_{3n}=\cdots=0$.
If $A_0=0$ for some $s$ (contrary to our assumption), then the remaining elements of $A$ become important
as $\varepsilon\to0$, and there are different cases to study; we do not consider these in this paper.

If $A_0\neq0$ makes a circuit $k$ times around the origin (say, as $s$ varies), then
following the eigenvalues $\lambda_\mu$ continuously along the circuit produces a net cyclic permutation $\mu\to \mu + k$ 
(${\rm mod}\,n$). This monodromy of the eigenvalues has the same form as that we mentioned \cite{kato}
in the first paragraph of Sec.\ \ref{intro}.  The situations are related
because $\varepsilon A$ can be considered as a perturbation from the exceptional point $H=J_n$ to an
``effective'' Hamiltonian $H'=J_n+\varepsilon A$, and we are considering a case in which the degeneracy
of eigenvalues is fully lifted. Henceforth we assume that $A_0$ does {\em not} encircle the origin
as $s$ varies from $0$ to $1$.

If, in the adiabatic limit $\varepsilon\to0$ and under our assumptions, a state prepared in the $\mu$th
instantaneous eigenvector of $J_n+\varepsilon A$ stayed in the corresponding eigenstate until $s=1$,
then there would be a change in its ``phase'' (i.e. the log of the amplitude, which here is complex) of
\be
\sum_{r=1}^n \varepsilon^{r/n-1}\zeta^{\mu r}\int_0^1 ds\, a_r(s)
\label{puisphase}
\ee
plus order $\varepsilon^{1/n}$, plus possibly a further contribution to the geometric phase, which we
discuss in Sec.\ \ref{final} below. Apart from the $r=n$ term, the terms displayed in expression 
(\ref{puisphase}) are {\em dynamical phases}, which depend on $T$. In addition to the usual one 
that is of order $T$ (absent here because we subtracted off the trace of $H$), there are also fractional 
powers of $T$ \cite{joye}. These may be  considered ``stretched exponential'' dependence on the 
time scale $T$ of adiabatic evolution.

In view of our choice that $\Lambda(1)=\Lambda(0)$, the final $r=n$ term in (\ref{puisphase}) is a
{\em geometric phase}, like the usual Berry phase in the case of non-degenerate eigenvalues, but given by
the average$\int ds\, n^{-1} {\rm tr}\, A$ of the diagonal elements of $A$. It can change
by a multiple of $2\pi i$ under a ``large'' residual gauge transformation that winds in ${\bf C}^\times$
as a function of $s$ (see Appendix \ref{app:gauge}); thus it is well-defined only modulo $2\pi i$. In other words, it is
the Berry phase {\em factor} or holonomy 
\be
e^{n^{-1}\int_0^1{\rm tr}\, A}
\ee 
that is well-defined if we do
not keep track of the choice of basis along the path (i.e.\ is fully gauge invariant). The dynamical
phase terms $r<n$ are gauge invariant (again, see Appendix \ref{app:gauge}); these gauge-invariance 
properties are similar to the usual Hermitian case.

%%%%%%%%%%%%%%%%%%%%%%%%%

\subsubsection{A special class of loops}

An important special case of adiabatic transport in $\cepn$ is that in which $A$ (and hence $a_r$) is
independent of $s$ (for all $r$). In that case, the coefficients in eq.\ (\ref{jplusepsa}) are constant,
so the system does stay in an initial eigenstate of $J_n+\varepsilon A$ for all $s$ if it is in one initially, 
and the preceding remarks conclude the calculation. We note that the corresponding path is ``straight'' 
in $GL_n({\bf C})$, with $\Lambda(s)=\Lambda(0)\exp (-sA)$,
and that such paths can return to the starting point, so $\Lambda(1)=\Lambda(0)$.
Moreover, these loops can be non-trivial in both $\pi_1(GL_n({\bf C}))$ and $\pi_1(\cepn)$, that is,
they can have non-zero winding number (modulo $n$) when projected to $\cepn$. In the general case, in
which $A$ is not constant, there could be further contributions to the geometric phase of the same order,
which we discuss next.

%%%%%%%%%%%%%%%%%%%%%%%%%%

\subsubsection{More general loops}
\label{final}

In order to examine the general scenario, we apply the adiabatic theorem to $J_n+\varepsilon A$. At leading
order, the $\mu$th eigenvector of $J_n+\varepsilon A$ can be chosen to be
\be
|v_\mu\rangle=\left(\begin{array}{c}
1\\
(\varepsilon A_0)^{1/n}\zeta^\mu\\
\cdot\\
\cdot\\
(\varepsilon A_0)^{(n-1)/n}\zeta^{\mu(n-1)}
\end{array}\right)[1+{\cal O}(\varepsilon^{1/n})],
\ee
which is periodic in $s$ under our assumptions.
If we use these instantaneous eigenvectors as a basis set (together with a dual basis as before),
then in this basis the evolution equation becomes
\be
\varepsilon\partial_s |v\rangle=(D+\varepsilon A')|v\rangle
\ee
where $D={\rm diag}\,(\lambda_0,\ldots,\lambda_{n-1})$, and $A'_{\mu\nu}=-\langle v_\mu|\partial_s
v_\nu\rangle$. Now we use the adiabatic theorem for this non-Hermitian non-degenerate
situation \cite{nenciu}. As mentioned already,
in this case, with eigenvalues $\lambda_\mu$ with
differences much larger than $\varepsilon$ as $\varepsilon\to0$, the adiabatic theorem in general does
not hold for all the eigenspaces of $D$, but only for the dominant mode (the one with the largest real
part of its eigenvalue). (In the less general case in which a permutation of the $|v_\mu\rangle$s makes
$A'$ block diagonal with the same block structure for all $s$, then the adiabatic theorem holds for the
dominant mode in each block.)
If we make the stronger assumption that $|\arg A_0(s)|<\pi$ for all $s$ (i.e.\ $A_0$ does not touch or
cross the negative real axis) then, for all $s$ and as $\varepsilon\to0$, ${\rm Re}\,\lambda_\mu$ is
largest when $\mu=0$, and is non-degenerate. [Technically, we assume that $A_0(s)$ does not approach
the negative real axis or zero closer than some small constant, say $\delta>0$.] With this assumption, it is not 
difficult to show that if the initial state is purely the dominant mode, then it remains in it for all $s$ with
sufficient accuracy as $\varepsilon\to0$ (see Ref.\ \cite{nenciu} and Appendix \ref{app:genadia}). 
Moreover, the additional contribution to the geometric or Berry ``phase''
is found by integrating the diagonal element $A'_{00}$ for the dominant mode.

In the present case, we find that
\be
A'_{\mu\nu}=-\sum_{r=1}^{n-1}\frac{r}{n^2}\zeta^{(\nu-\mu)r}\frac{\partial_sA_0}{A_0}+{\cal
O}(\varepsilon^{1/n}),
\ee
so the diagonal elements are given by $\frac{1-n}{2n}\partial_s\ln A_0$ to leading order.
Integrating from $s=0$ to $1$,
the change in the complex amplitude is $[A_0(0)/A_0(1)]^{(n-1)/(2n)}$. Because $A_0$ does not make a
circuit around the origin, this factor is $1$. Then the net phase change through order
$\varepsilon^0$ is given by the integrated Puiseux expansion (\ref{puisphase}). It is remarkable that,
for adiabatic transport in $\cepn$, there is an order $\varepsilon^0$ part of $A'$, yet this part still
does not contribute to the geometric phase.

%%%%%%%%%%%%%%%%%%%%%%%
\subsubsection{Calculation of net holonomy}
\label{holonomy}

Finally, we need to calculate the net geometric phase for a loop. First, because $\Lambda(1)=\Lambda(0)$, the
non-Abelian holonomy [which is an element of $GL_n({\bf C})$] is ${\cal P}\exp \int_0^1 A\, ds$
(${\cal P}\exp$ is path-ordering of the exponential). Hence the holonomy for any loop is
$\Lambda(1)^{-1}\Lambda(0)=I$, and by considering a
contractible loop and using Stokes's theorem, it follows that the non-Abelian Berry connection $A$ has zero
(Yang-Mills) curvature. 

The trace of $A$ is in the Lie subalgebra $gl_1({\bf C})\cong {\bf C}$ of $gl_n
({\bf C})$, so the corresponding Abelian holonomy is $\det \Lambda(0)/\det\Lambda(1)=1$ in $GL_1({\bf
C})\cong{\bf C}^\times$. Then $n^{-1}{\rm tr}\,A$, which determines the geometric phase, is again a
flat connection (i.e.\ its Berry curvature is zero). Its holonomy is
\be
e^{n^{-1}\! \int_0^1{\rm tr}\,A\,ds}=[\det \Lambda(0)/\det\Lambda(1)]^{1/n}
\label{eq:holon}
\ee
[with the $n$th root defined by imposing continuity on $\{\det\Lambda(s)\}^{1/n}$];  because of 
the vanishing curvature, it depends only on the homotopy class of the loop. For any contractible loop in 
$GL_n({\bf C})$ this holonomy is $1$, but for a non-contractible loop it is an $n$th root of $1$, and so 
a power of $\zeta$. [Note that $\pi_1(GL_n({\bf C}))\cong {\bf Z}$, so such
loops exist; they are associated with the non-contractibility of ${\bf C}^\times$.]
It is precisely $\zeta^{-1}$ to the power of the winding number of the path of $\det \Lambda(s)$ around the
origin in ${\bf C}^\times$; note that this winding number changes by a multiple of $n$ under a residual
gauge transformation. In $\cepn$, the fundamental group is ${\bf Z}_n$, so repeating a given loop $n$
times produces a loop that is contractible in $\cepn$, and hence the $n$th power of the holonomy for any loop
must be $1$, consistent with our conclusion that the holonomy is a power of $\zeta$. 
This concludes the derivation of the general results. In Appendix \ref{app:geom}, we also explain that 
the mapping from a loop in $\cepn$ to a power of $\zeta$, given by the holonomy, Eq.\ (\ref{eq:holon}),
can be viewed as a torsion first Chern class $c_1$.

%%%%%%%%%%%%%%%%%%%%%%%%

\subsubsection{Example: $n=2$}

As an illustration of the general results, we solve the $n=2$ model explicitly for $A$ constant and $A_0
\neq 0$. An example of a non-contractible loop in $\mathcal{EP}_2$ is parametrized by $\textbf{X}(s) =
( i \cos\phi(s), i \sin\phi(s), 1)^T$ with $\phi(s) = 2\pi s$. Then with $|u_1\rangle=(1,ie^{i\phi})^T$ and
$|u_2\rangle=(1,0)^T$, we find
\be
A=\left(\begin{array}{cc}-2\pi i&0\\
                         2\pi i&0
                            \end{array}\right),
\ee
so $A_0=2\pi i$, and the eigenvalues of $J_2+\varepsilon A$ are
\bea
\lambda_{0,1} &=&{} -i \pi \varepsilon \pm \sqrt{ 2\pi i \varepsilon - \pi^2 \varepsilon^2 } \\
&=& {}\pm \pi^{1/2} (1+i) \varepsilon^{1/2} - i \pi \varepsilon + \mathcal{O}( \varepsilon^{3/2} ). 
\label{twoEV}
\eea
The two leading terms agree with $a_1=A_0^{1/2}$ and $a_2=\frac{1}{2}{\rm tr}\, A=-i\pi$, and further
$a_{2k}=0$ for $k>1$; note that the Berry phase is $\pi$ (modulo $2\pi$). Our general theory tells us
that these results are independent of the choice of gauge, and that the Berry phase is invariant under
sufficiently slowly varying smooth changes in the loop. More generally, if we
choose $\phi(s)=2\pi m s$ for an integer $m$, giving a loop of winding number $m$, then the holonomy 
is $(-1)^m$.

%%%%%%%%%%%%%%%%%%%%%%%%%%%%%%%%%%%%%%%%%%%%%%%%%%

\section{Experimental implementations for $n=2$}
\label{expt}

The simplest demonstration of our results 
is for non-Hermitian $2\times 2$ Hamiltonians, which are of the general form in Eq.\ \eqref{twoham}. Here, 
we discuss two experimental setups which realize such a Hamiltonian, and we propose noncontractible loops 
along which adiabatic transport would produce the results in Eq.\ \eqref{twoEV}. In either setup, this means 
that if the system were initialized in the dominant mode, there would be a contribution to the (complex) 
dynamical phase of $(1+i) \sqrt{\pi T}$, and the Berry phase factor would be $-1$. 

The first setup consists of a qubit (two-level system) that is coupled to a waveguide, and which can decay to 
a third level. When the system is post-selected for evolution that remains within the qubit's Hilbert space, the 
resulting dynamics can be described via an effective Hamiltonian that is non-Hermitian, as demonstrated 
experimentally in Ref.\ \cite{kater}.  In the rotating frame and rotating wave approximation, we identify 
the three complex numbers 
$X$, $Y$, and $Z$ in Eq.\ \eqref{twoham} with experimental parameters: $X=J \cos(\phi)$, $Y=J \sin(\phi)$,  
$Z = \Delta/2 - i \gamma/4$, where $\Delta$ is the detuning of the drive applied to the qubit (via 
the waveguide), $\gamma$ the difference in decay rates of the two qubit states, $J$ the coupling 
strength (also known as 
Rabi frequency), and $\phi$ the (Rabi) phase. From the discussion below Eq.\ \eqref{twoham}, we infer that 
exceptional points lie at $\Delta = 0$ and $J=|\gamma|/4$ for all $\alpha$. For fixed $\Delta = 0$ and 
$J=|\gamma|/4$, varying $\phi$ by $2\pi$ describes a noncontractible loop. If $\gamma$ is also controlled
\cite{kater}, then the accessible subspace of $\mathcal{EP}_2$ forms a frustum---a cone without its apex 
(because there, $H=0$ and so is diagonalizable). 

Second, we argue that full control over Eq.\ \eqref{twoham} can be achieved with the
optomechanical device used in Ref.\ \cite{Xu2018}. The device consists of a dielectric membrane in the middle 
of an optical cavity. The membrane's vibrational modes can be coupled to each other by 
sending light at particular frequencies into the cavity; in addition to the complex-valued mutual coupling, light 
also introduces a complex-valued self-coupling term to each oscillating normal mode \cite{Aspelmeyer2014}. 
In Ref.\ \cite{Xu2018}, two pairs of light beams were used to couple 
two oscillating normal modes. It is straightforward to show that various combinations of laser tones (i.e.\
their power and detuning) can be chosen to give independent control over the complex parameters 
$X$, $Y$, and $Z$ in Eq.\ (\ref{twoham}). 

%%%%%%%%%%%%%%%%%%%%%%%%%%%

\section{Conclusion}
\label{conclusion}

To conclude, we studied adiabatic transport around a loop in a space of exceptional points of type $EP_n$
for $n\times n$ Hamiltonian matrices.
We found that the dynamical phase is given by a Puiseux series of fractional powers of $T$, and that the
Berry phase (modulo $2\pi$) depends only on the homotopy class of the loop. The results hold for a
choice of initial state that depends on both the loop (which must be in a certain class of loops), and on
$T$.

Clearly, it would be of interest to carry out an analysis by similar methods in other cases, such as when $A_0=0$ all along
the loop. Alternatively, for $n>2$ we can also consider, for example, spaces of EPs of types $EP_{n'}$
for $n'<n$ within $M_n({\bf C})$, or spaces of Hamiltonians for which the Jordan canonical form is a
direct sum of several
Jordan blocks with the same eigenvalue. We expect these cases to involve the non-Abelian connection for
transporting a proper subspace \cite{wz}, as well as effects similar to those we found for a Jordan block
of size $n$. We leave these cases for later study.

\acknowledgements

We are grateful for discussions with A. Alexandradinata and G. Moore.
We acknowledge support from Yale University (JH), from AFOSR grant FA9550-15-1-0270 and ONR MURI
N00014-15-1-2761 (JGEH), and (while the paper was under revision) from NSF grant no.\ DMR-1724923 
(JH and NR).

%%%%%%%%%%%%%%%%%%%%%%%%%%%%%%%%%%%%%%%%
\appendix

%%%%%%%%%%%%%%%%%%%%%%%%%%%%%%%%%%%%%%%%%%
\section{Geometry and topology of $\cepn$}
\label{app:geom}

In this Appendix, we consider the geometry and topology of the space $\cepn$. We first show that it is a
complex manifold of complex dimension $n(n-1)$, and has the same topology (i.e.\ what is called homotopy type
\cite{hatcher}) as $SU(n)/{\bf
Z}_n$ [for $n=2$, this becomes $SO(3)$, the space of real $3\times 3$ orthogonal matrices of determinant $1$].
This space is connected but not simply connected; its fundamental group is $\pi_1(\cepn)\cong {\bf Z}_n$, which
is not the trivial group when $n\geq 2$, and corresponds to the $n$-fold connectedness described in Sec.\ 
\ref{outline}. (The calculation in the main text is required in order to determine the Berry phase factor, relate it to a 
holonomy, show that there are loops for which it is not unity, and relate it to the winding number of the loop.) 
Finally, we also determine the low-dimensional 
homology and cohomology groups of $\cepn$ with integer coefficients, including the second cohomology 
group $H^2(\cepn)$, which is again $\cong {\bf Z}_n$. The latter group tells us about the possible holonomy 
for a loop in $\cepn$, and we explain how the results in the text can be viewed as an example of a torsion first 
Chern class, which is an element of $H^2(\cepn)$.

To begin, for $H=\Lambda J_n\Lambda^{-1}$ and $\Lambda$ an invertible complex matrix [an element of the
general linear group $GL_n({\bf C})$], we see that multiplying $\Lambda$ on the right by any
invertible matrix that commutes with $J_n$ produces the same $H$. The latter matrices form the 
subgroup ${\bf C}^\times\times{\cal J}_n$ defined in Sec.\ \ref{geom}. It follows that the space of such 
$H$ is $\cepn\cong GL_n({\bf C})/[{\bf C}^\times \times {\cal J}_n]$, which is thus a homogeneous complex 
manifold. The complex dimension of $GL_n({\bf C})$ is $n^2$, and that of ${\bf C}^\times \times
{\cal J}_n$ is $n$, so the complex dimension of $\cepn$ is $n(n-1)$. This number can also be understood
heuristically as follows: $M_n({\bf C})$ has dimension $n^2$, and a generic matrix has $n$ distinct
complex eigenvalues. The condition that all eigenvalues be zero imposes $n$ constraints, leaving an
$n^2-n$ dimensional space. This space contains points not in $\cepn$, but it turns out that those points
form subspaces of dimension $<n^2-n$; for example, for any $n\geq 2$ the space includes the zero matrix,
which is not similar to any non-zero matrix.

To study further the geometry and topology of $\cepn$, a useful first step is to analyze that of
$GL_n({\bf C})$. (The result of the division of $GL_n({\bf C})$ by the product group can then be
investigated afterward, one factor at a time.) If we arbitrarily choose an inner product on ${\bf C}^n$,
say the standard one, then any
invertible matrix $g\in GL_n({\bf C})$ can be expressed in the polar decomposition $g=Uh$, where $U$ is
unitary and $h$ is a positive-definite Hermitian matrix (i.e.\ all its eigenvalues are strictly positive).
This can be obtained from $h=(g^\dagger g)^{1/2}$, where the positive square root is taken for each
positive eigenvalue of $g^\dagger g$, and $U=gh^{-1}$. The space of such $h$ is contractible; $h$ can be
deformed to the identity $I$. Hence $GL_n({\bf C})$ has a deformation retract \cite{hatcher} to the
space $U(n)$ of unitary matrices.

For the second step, the group $GL_n({\bf C})/{\bf C}^\times$ [where ${\bf C}^\times$ is embedded in
$GL_n({\bf C})$ as the subgroup of non-zero complex multiples of the identity] is also known as the
projective linear group $PGL_n({\bf C})$. It has a deformation retract onto $U(n)/U(1)\cong SU(n)/{\bf
Z}_n$. Here ${\bf Z}_n\subset U(1)\subset U(n)$ is embedded in $SU(n)$ as the subgroup of $n \times n$ 
matrices that are a power of $\zeta$ times the identity; it is the center of $SU(n)$. (Note that, throughout 
the paper, we use $\cong$ to stand for topological isomorphism [homeomorphism] of topological spaces; 
when the space is also a group, the map
is also an isomorphism of groups. In the special case of a discrete group, the discrete topology is used.)

Finally, we must also mod out by ${\cal J}_n$. It is a contractible subgroup of $GL_n({\bf C})$, and
intersects ${\bf C}^\times I$ only at $I$, so its image is still a contractible subgroup in $PGL_n({\bf
C})$. As for any space that is the quotient space $G/H$ of a group $G$ by a subgroup $H\subseteq G$,
$PGL_n({\bf C})$ can be viewed as a fibre bundle \cite{hatcher} over the quotient space $\cepn$ with fibre $\cong
\cal J$. Because the fibre is contractible, $\cepn$ has the same homotopy type as $PGL_n({\bf C})$, or
as $SU(n)/{\bf Z}_n$. That is, these spaces are homotopy equivalent, due to the existence of deformation
retracts from $PGL_n({\bf C})$ to $\cepn$, and from either of these to $SU(n)/{\bf Z}_n$.

The homotopy type of a space can also be studied by using the homotopy groups. We can apply the homotopy
long exact sequence of a fibration to the fibre bundle $G$ over $G/H$ with fibre $H$ (where $H\subseteq
G$ are groups) \cite{hatcher},
\be
\cdots\to\pi_{i+1}(G)\to\pi_{i+1}(G/H)\to\pi_i(H)\to\pi_i(G)\to\cdots,
\ee
which holds down to $\pi_0(G/H)$ [for $i=0$, $\pi_0$ of a space is in general a set without a group
structure, however for the present case the $\pi_0$s are groups, except for $\pi_0(G/H)$ in the case that
$H$ is not a normal subgroup of $G$]. If we apply this with $G=PGL_n({\bf C})$ and $H={\cal J}_n$, then as
all homotopy groups (or sets) of ${\cal J}_n$ are zero (because it is contractible), we of course find again
that $\pi_i(\cepn)\cong \pi_i(PGL_n({\bf C}))\cong\pi_i(SU(n)/{\bf Z}_n)$ for all $i\geq0$. Applying the
sequence again with $G=SU(n)$, $H={\bf Z}_n$, and using (with $n\geq 2$ from here on)
$\pi_1(SU(n))=\pi_2(SU(n))=0$, $\pi_3(SU(n))\cong{\bf Z}$, and of course $\pi_i({\bf Z}_n)=0$ for $i>0$,
$\pi_0({\bf Z}_n)={\bf Z}_n$, we obtain $\pi_1(\cepn)={\bf Z}_n$, $\pi_2(\cepn)=0$, $\pi_3(\cepn)={\bf Z}$.
For $n=2$, these results are well-known for ${\cal EP}_2\cong{\bf RP}^3\times {\bf R}\cong SO(3)\times{\bf
R}$, and for general $n$ they are fairly well-known for $SU(n)/{\bf Z}_n$. We note that, when the fibre 
is a discrete group, these results can also be understood in terms of covering spaces; for example, $SU(n)$ 
is the simply-connected covering space of $SU(n)/{\bf Z}_n$, which means that 
$\pi_1(SU(n)/{\bf Z}_n)\cong {\bf Z}_n$.

Next we turn briefly to the homology and cohomology of $\cepn$, which depend only on the
homotopy type of the space, and their applications (these are not essential for understanding the main
text). The first homology group, with integer coefficients, is $H_1(\cepn)\cong \pi_1(\cepn)\cong
{\bf Z}_n$ as $\pi_1(\cepn)$ is Abelian. By the universal coefficient theorem \cite{hatcher}, the
cohomology with integer
coefficients is $H^1(\cepn)=0$, and $H^2(\cepn)\cong F_2\oplus T_1$, where $F_2$ is the free part (group
of elements of infinite order) of $H_2(\cepn)$, $T_1$ is the torsion part (group of elements of finite
order) of $H_1(\cepn)$, and the isomorphism is non-canonical. Thus $H^2(\cepn)$ contains a subgroup $\cong
{\bf Z}_n$. For $n=2$, $H^2({\bf RP}^3)\cong{\bf Z}_2$, while $H^3({\bf RP}^3)\cong{\bf Z}$. For $n\geq 2$,
using theorems of Hopf (Ref.\ \cite{brown}, pp.\ 1--2, 41--42), there is a surjection from $\pi_2(\cepn)=0$
onto $H_2(\cepn)$, because the {\em group} homology $H_2(\pi_1(\cepn))=H_2({\bf Z}_n)=0$, and so
$H_2(\cepn)=F_2=0$, implying $H^2(\cepn)\cong{\bf Z}_n$. Alternatively, we can use a Cartan-Leray
spectral sequence (see Ref.\ \cite{brown}, p.\ 173 or Ref.\ \cite{weibel}, Sec.\ 6.10) to obtain
$H^2(\cepn)\cong {\bf Z}_n$.

In general, the elements of the second integral cohomology group $H^2(X)$ for a topological space $X$ 
correspond one-to-one with the isomorphism classes of complex rank one vector bundles  (also called 
line bundles) over $X$; the group operation corresponds to taking tensor products of line bundles. 
The element in $H^2(X)$ for a given line bundle is the first 
Chern class $c_1$ of the bundle; thus $c_1$ completely classifies line bundles up to isomorphism. 

Our adiabatic transport construction does not directly produce 
a line bundle (because e.g.\ the dominant state depends on the path used), but the $gl_1({\bf C})$ connection 
$n^{-1}{\rm tr}\, A$, or more precisely its holonomy around all possible loops, uniquely determines a line bundle 
(up to isomorphism), and hence determines its first Chern class $c_1 \in
H^2(\cepn)\cong {\bf Z}_n$. (Our connection is flat---see  Sec.\ \ref{holonomy}---but the statement would hold 
even if the connection were not flat; see e.g.\ Ref.\ \cite{fms}, Sec.\ 2, for exposition and references.) 
Because our connection is flat, our formula Eq.\ (\ref{eq:holon}) for the holonomy is precisely this $c_1$; 
when given a homotopy class of loops [in $\pi_1(\cepn)$], or the corresponding homology class of cycles 
[in $H_1(\cepn)$], it specifies an element of ${\bf Z}_n$, represented multiplicatively as a phase factor, 
which is the holonomy.
Viewed as an element of $H^2(\cepn)$ (additively), $c_1$ is non-zero and in fact is a generator of $H^2(\cepn)$. This 
$c_1$  is an example of a torsion first Chern class, that is, one that is not equivalent to an integer or set of integers.

%%%%%%%%%%%%%%%%%%%%%%%%%%%%%%%%%%%%%%%%%%%%

\section{Gauge transformations and invariance of results}
\label{app:gauge}

The evolution equation (\ref{jplusepsa}) is covariant under $s$-dependent transformations lying in the subgroup
${\bf C}^\times \times {\cal J}_n$. Precisely, if the column vector $|u\rangle$ is replaced by
$|u^g\rangle=g|u\rangle$, where $g\in{\bf C}^\times \times {\cal J}_n$, then the evolution equation becomes
$\varepsilon\partial_s|u^g\rangle=(J+\varepsilon A^g)|u^g\rangle$, where
\be
A^g=gAg^{-1}+\partial_s g.g^{-1}.
\ee
As $g$ is upper triangular, the inhomogeneous term $\partial_s g.g^{-1}$ is as well (it is in the Lie
algebra of ${\bf C}^\times \times {\cal J}_n$).

We show in Appendix \ref{app:evalscal} below that the leading terms $\propto \varepsilon$ in the
coefficients $c_i$ in the characteristic equation contain elements of $A$ on or below the
diagonal. It follows that the inhomogeneous terms in $A^g$ have no effect on the coefficients $c_r$
in order $\varepsilon$ except for the diagonal of $A^g$ which affects $c_{n-1}\propto {\rm tr}\,A$.
Moreover, $J_n+\varepsilon gAg^{-1}=g(J_n+\varepsilon A)g^{-1}$ has the same eigenvalues as
$J_n+\varepsilon A$. Hence the terms displayed in expression (\ref{puisphase}) are gauge invariant to the
order shown, except for the $r=n$ term; the latter transforms as a ${\bf C}^\times$ connection. Then the
$s$ integrals of these terms are also gauge invariant, except that the Berry phase ($r=n$) term changes
by a multiple of $2\pi i$ (where here $i$ is the square root of $-1$, not an index), and thus is invariant 
except under a ``large'' gauge transformation, that is one that winds around the origin in ${\bf C}^\times$.

%%%%%%%%%%%%%%%%%%%%%%%%%%%%%%%%%%%%%%%%%%

\section{Scaling of the characteristic equation and eigenvalues}
\label{app:evalscal}

The characteristic equation of $J_n+\varepsilon A$ has the form
\be
\lambda^n + \sum_{i=0}^{n-1}c_i\lambda^i=0
\ee
where the $c_i$ are similarity invariants of $J_n+\varepsilon A$, and all are of order ${\cal O}
(\varepsilon)$ as $\varepsilon\to0$. Indeed, the terms of first order in $\varepsilon$ in $c_i$ are
$c_i=-\varepsilon \sum_{j=0}^i A_{n+j-i,1+j}+{\cal O}(\varepsilon^2)$ for all $i=0$, \ldots, $n-1$.

Because any root of the equation must tend to zero as $\varepsilon\to0$, it is not difficult to see that
the $c_0$ term is the most important of the terms containing a $c_i$, provided that
$\lim_{\varepsilon\to0}c_0/\varepsilon\neq 0$. Then $\lambda\sim (-c_0)^{1/n}\propto \varepsilon^{1/n}$,
and the other $c_i$ ($i\neq0$) do not contribute at leading order in this limit.

%%%%%%%%%%%%%%%%%%%%%%%%%%%%%%%%%%%%%%%%%%%%%%

\section{Generalized adiabatic theorem}
\label{app:genadia}

Here for completeness we prove the generalized version of the adiabatic theorem (including the Berry phase)
in the context of the main text. The result is contained in Ref.\ \cite{nenciu}, but our proof is
different. We consider the evolution equation
\be
\partial_s|v\rangle=(\varepsilon^{-1}D+A')|v\rangle,
\ee
where $D={\rm diag}\,(\lambda_0,\ldots,\lambda_{n-1})$ and $A'$ is the Berry connection. In our case,
$D$ consists of eigenvalues proportional to $\varepsilon^{1/n}$ that are never equal, and $A'$ has entries
independent of $\varepsilon$. We assume that $\lambda_0$ has largest real part (i.e.\ it is dominant), and that
the differences of the real parts of the $\lambda_\mu$s are bounded away from zero (this holds under the 
assumptions in the text).
We also assume that all elements of $A'$ are bounded in magnitude by the same constant $B>0$ uniformly
for all $s$. First, we suppose that the system is prepared in the $\mu=0$ (dominant) eigenstate at $s=0$,
and consider the amplitude for it to return to that eigenstate at $s=1$. If the possible transitions to
other modes ($\mu\neq0$) are neglected, then the change in the complex amplitude of the dominant state
will be a factor $\exp \{\int_0^1 ds\,[\varepsilon^{-1} \lambda_0(s)+A'_{00}(s)]\}$. We will show that
corrections
to this due to transitions, and the amplitude for ending in a different state, are of relative size ${\cal
O}(\varepsilon^{[n-1]/n}$) at most. We emphasize that our general argument applies whenever
$D$ is diagonal and the differences of the real parts of the diagonal entries from the
dominant one are bounded below by a non-zero constant times $\varepsilon$ to any power $<1$.

We first extract the factor $\exp \{\int_0^1 ds\,[\varepsilon^{-1} \lambda_0(s)+A'_{00}(s)]\}$, to calculate
relative
to this expected factor; this has the effect of taking $\lambda_0=0$ (and $A_{00}'=0$) without loss of
generality, by subtracting these from the diagonal elements of $D$ (resp.,\ $A'$). Now we begin by
considering $n=2$, and set $A_{11}'=0$ for now. The change in amplitude is
\be
\langle v_0|{\cal P}\exp \int_0^1 ds\,[\varepsilon^{-1} D + A']|v_0\rangle,
\ee
where the initial $|v_0\rangle=(1,0)^T$ is the dominant eigenstate of $J_n+\varepsilon A$ in this basis
(and $\langle v_0|$ is the corresponding element of the dual basis). This transition amplitude is
\bea
&=& \sum_{r=0}^\infty\int_{\cal D}\prod_{j=1}^{2r} ds_j\,A_{01}'(s_{2r})A_{10}'(s_{2r-1})\cdots
A_{10}'(s_1)\non\\
&&{}\times e^{\int_{s_1}^{s_2}\varepsilon^{-1}\lambda_1(s')\,ds'+\cdots +
\int_{s_{2r-1}}^{s_{2r}}\varepsilon^{-1}\lambda_1(s')\,ds'},
\eea
where ${\cal D}$ is the region defined by $s_1<s_2<\cdots <s_{2r}$ and all $s_j$ in $[0,1]$. Subtracting
the $r=0$ term (which is unity), and taking the absolute value, we have the bound
\be
\leq \sum_{r=1}^\infty\int_{\cal D}\prod_{j=1}^{2r} ds_j\,B^{2r}
e^{-\varepsilon^{-1/2} L\sum_{j=1}^r (s_{2j}-s_{2j-1})},
\ee
where $\varepsilon^{1/2} L>0$ is a lower bound on $-{\rm Re}\,\lambda_1>0$ for all $s\in[0,1]$, which
exists by our assumptions. [This expression, with $1$ added, can be viewed as the partition function of
a statistical mechanics problem of domain walls at positions $s_j$, where at $s=0$, $1$, the state vector is
fixed at $|v_0\rangle$, and transitioning to the other state $|v_1\rangle=(0,1)^T$ carries a fugacity $B$ 
and energy penalty
$\varepsilon^{-1/2} L>0$ per unit length. As this penalty is large, domain walls are bound in pairs, and
it is unlikely that state $|v_1\rangle$ is found.] This is in turn less than or equal to
\be
\leq \sum_{r=1}^\infty\frac{1}{r!}\int_{{\cal D}'}\prod_{j=1}^{2r} ds_j\,B^{2r}
e^{-\varepsilon^{-1/2} L\sum_{j=1}^r (s_{2j}-s_{2j-1})}
\ee
(where ${\cal D}'$ is the domain $s_1<s_2$, $s_3<s_4$, \ldots, $s_{2r-1}<s_{2r}$  and all $s_j$ in
$[0,1]$), because the parts where some of the intervals $[s_{2j-1},s_{2j}]$ overlap give positive
contributions, and discarding these leaves a region that covers $\cal D$ $r!$ times. This multiple
integral is a product, and gives rise to an exponential series with the initial term $1$ omitted. Each
two-dimensional integral factor can be evaluated to give $\varepsilon^{1/2}/L + \varepsilon
(e^{-\varepsilon^{-1/2}L}-1)/L^2$, where the subleading
terms are introduced by the integration limits at $s=0$, $1$. Hence we have found the upper bound
\be
=\exp [\varepsilon^{1/2} B^2/L + \varepsilon B^2(e^{-\varepsilon^{-1/2}L}-1)/L^2] - 1.
\ee
As $\varepsilon\to0$, this gives $\sim \varepsilon^{1/2}B^2/L$, which is simply the first term in the
series.

In general, one should include $A_{11}'$. This can be added to $\varepsilon^{-1/2}\lambda_1$, and can be
absorbed into a change in the bound $L$ when $\varepsilon$ is sufficiently small. Hence the full result is
\bea
\lefteqn{\langle v_0|{\cal P}\exp \int_0^1 ds\,[\varepsilon^{-1} D + A']|v_0\rangle=}&&\\
&&\qquad\exp \left\{\int_0^1 ds\,[\varepsilon^{-1} \lambda_0(s)+A'_{00}(s)]\right\}\cdot[1+{\cal
O}(\varepsilon^{1/2})]\qquad\non
\eea
for the $n=2$ case. Similarly, we can show that the amplitude for making a transition to the state $|v_1\rangle$ 
during the evolution is
\bea
\lefteqn{\langle v_1|{\cal P}\exp \int_0^1 ds\,[\varepsilon^{-1} D + A']|v_0\rangle=}&&\\
&&\qquad{\cal
O}(\varepsilon^{1/2}B/L)\cdot\exp \left\{\int_0^1 ds\,[\varepsilon^{-1}
\lambda_0(s)+A'_{00}(s)]\right\}\qquad\non
\eea
as $\varepsilon\to0$, and the same bound also applies to the amplitude for starting in $1$ and ending
in $0$.

Finally, for $n>2$ we can absorb $A'_{\mu\mu}$ ($\mu>0$) into a change in the lower bound on all $-{\rm
Re}\,\lambda_\mu$, $\mu\neq0$. The remaining elements of $A'_{\mu\nu}$ with $\mu$, $\nu\neq 0$, which
produce transitions among those modes, are bounded in magnitude by $B$, and in the above argument simply
lead to another order $1$ contribution that can also be absorbed into $\varepsilon^{(1-n)/n}L$. Hence the
result is similar, with corrections smaller at least by $\sim\varepsilon^{(n-1)/n}B^2/L$, and similarly for 
the amplitude for a transition to any state $\mu\neq0$.

%%%%%%%%%%%%%%%%%%%%%%%%%%%%%%%%%%%%%%


\begin{references}
\bibitem{Xu2018} Such subspaces can be accessed if there is complete control over the matrix elements
of $H$, which can be achieved for $n=2$ in the optomechanical device demonstrated in H. Xu, L. Jiang,
A.A. Clerk, and J.G.E. Harris, Nature {\bf 568}, 65 (2019).
\end{references}
\end{document}